\documentstyle[12pt]{article}
\begin{document}
\newcommand{\mup}{{\mu}^{\prime}}
\newcommand{\nup}{{\nu}^{\prime}}
\newcommand{\muz}{{\mu}^{\prime \prime}}
\newcommand{\nuz}{{\nu}^{\prime \prime}}
\newcommand{\z}{\prime \prime}
\newcommand{\p}{\prime}
\newcommand{\rond}{\partial}
\newcommand{\bq}{\textbf{q}}
\newcommand{\bp}{\textbf{p}}
\newcommand{\bz}{\textbf{0}}
\begin{titlepage}
\vspace{-10mm}

\begin{center}
\begin{large}
{\bf THE DIFFICULTY OF \\ SYMPLECTIC ANALYSIS \\ WITH SECOND CLASS
SYSTEMS   }\\
\end{large}
\vspace{5mm}
 {\bf A. Shirzad \footnote{e-mail:
shirzad@ipm.ir}},
 {\bf M. Mojiri \footnote{e-mail:
mojiri@sepahan.iut.ac.ir}}

%\vspace{12pt}\\
{\it Department of  Physics, Isfahan University of Technology \\
Isfahan,  IRAN, \\
Institute for Studies in Theoretical Physics and Mathematics\\
P. O. Box: 5531, Tehran, 19395, IRAN.} \vspace{0.3cm}
\end{center}
\abstract{ Using the basic concepts of chain by chain method we
show that the symplectic analysis, which was claimed to be
equivalent to the usual Dirac method, fails when second class
constraints are present. We propose a modification in symplectic
analysis that solves the problem.
 \vfill}
\end{titlepage}
\section{Introduction}
There are some attempts to study a constrained system in the
framework of  first order Lagrangian \cite{1,2}. The coordinates
appearing in a first order Lagrangian are in fact the phase space
coordinates. The Euler-Lagrange equations of motion of a first
order Lagrangian in an ordinary (non-constrained) system are the
same as the canonical equations of motion. The kinetic term in a
first order Lagrangian constitutes of a one-form whose exterior
derivative appears in the equations of motion. The resulted
two-form, called the symplectic tensor, is singular for a
constrained system. If the system is not constrained, usually the
inverse of the symplectic tensor exists and provides the
fundamental Poisson brackets (we exclude degenerate systems
discussed in \cite{SeaRicZan,MisZan} in which the symplectic
tensor may have a lower rank in some regions of the phase space).

The properties of a constrained system can be determined by trying
to overcome the singularity of the symplectic tensor. Faddeev and
Jackiw \cite{FJ} used the Darboux theorem to separate canonical
and non-canonical coordinates.They solved the equations of motion
for non-canonical coordinates either  to decrease the degrees of
singularity of  the symplectic tensor or to find the next level
constraints.

Then using a special system of coordinates, the authors of
\cite{GP} showed that the Faddeev-Jackiw approach is essentially
equivalent to the usual Dirac method \cite{Dirac}. In a parallel
approach, known as symplectic analysis \cite{Sym1,Sym2,Sym3,MONTA}
one extends the phase space to include the Lagrange multipliers.
In this approach the consistency of constraints at each level adds
some additional elements to the symplectic tensor. In other words,
the kinetic part of the (first order) Lagrangian is responsible to
impose the consistency conditions.

The important point in most papers written in Faddeev-Jackiw
method or symplectic analysis is that they often show their
results for the constraints in the first level and then
\textit{deduce} that the same thing would be repeated at any
level. However, the whole procedure of studying the singularities
of  symplectic tensor, demonstrates some global aspects. For
example, some questions that may arise are as follows:

What happens, after all, to the symplectic tensor? Is it
ultimately singular? How many degrees of singularity may it have?
What is the relation of ultimate singularities with the gauge
symmetries of the system? and so on. In \cite{shimoj} we showed
that the symplectic analysis gives, at each step, the same results
as the traditional Dirac method (in the framework of \textit{level
by level approach}). The symplectic analysis may also be studied
 in the framework of \textit{chain by chain approach}\cite{lorshir}
  to obtain the Dirac constraints.

Meanwhile, some recent observation \cite{rothe} shows that in some
examples the result of symplectic analysis and the
well-established method of Dirac are not the same. This creates
serious doubt about the validity of the symplectic analysis.
Therefore, it is worth studying the origin of the difference
between this approach and that of Dirac \cite{Dirac}. This is the
aim of this paper. In the next section we first review the basic
concept of symplectic approach as given in \cite{shimoj}. As we
will show the symplectic analysis is equivalent to a special
procedure in Dirac approach in which one uses the extended
Hamiltonian at each level of consistency. In section(3) we will
show that in the framework of Dirac method one is not allowed to
use an extended Hamiltonian when there exist second class
constraints. The important point to be emphasized is that this
result can be understood more clearly in the framework of chain by
chain method. In section(4) we show that for a one chain system
with second class constraints the symplectic analysis as proposed
in the literature fails. This result can be simply generalized to
the general case of a multi-chain system. When recognizing the
origin of the problem, we give our prescription to solve it in
section(5). Finally in section(6) we give an example.

The last point to be noticed is that the problem would  not show
itself for systems with two levels of constraints. As we will
show, this is the case for second class systems with at least four
levels of constraints. That is the reason for the fact that the
problem does not appear if one considers just first level of
constraints.

\section{Review of symplectic approach}
Consider a phase space with coordinates $y^i(i=1 , \dots ,2K)$
specified by the first order Lagrangian
\begin{equation}
L= a_i(y) \dot{y}^i-H(y) \label{a1}
\end{equation}
where $H(y)$ is the canonical Hamiltonian of the system. The
equations of motion read
\begin{equation}
f_{ij}\dot{y^j}=\partial_i H \label{a2}
\end{equation}
where $\rond_i \equiv \frac{\rond}{\rond y^i}$ and the
presymplectic tensor $f_{ij}$ is defined as
\begin{equation}
f_{ij} \equiv \partial_i a_j(y) - \partial_j a_i(y). \label{a3}
\end{equation}
We denote it in matrix notation as $f$. This matrix is invertible
for a regular system. Let $f^{ij}$ be the components of the
inverse, $f^{-1}$. From (\ref{a2}) we have
\begin{equation}
\dot{y}^i=\left\{ y^i,H \right\} \label{a3.5},
\end{equation}
where the Poisson bracket $\left\{\ ,\ {}\right\}$ is defined as
\begin{equation}
\left\{ F(y),G(y)\right\}=f^{ij} \partial_i F\partial_j G .
 \label{a4}
\end{equation}
If $f$ is singular, then using the Darboux theorem, as shown in
\cite{FJ}, one can choose the independent coordinates $(y^{\p
\alpha},\lambda^l )$ such that
\begin{equation}
L=a^{\p}_{\alpha}\dot{y}^{\p \alpha}-\lambda^l \Phi_l (y^{\p})
-H(y^{\p}) \label{s1}
\end{equation}
where $f^{\p}_{\alpha \beta}= \rond _{\alpha}a^{\p}_{\beta}-\rond
_{\beta}a^{\p}_{\alpha}$ is invertible. This shows that one can
consider a system with a singular tensor $f_{ij}$, as a regular
one described by
\begin{equation}
L=a^{\p}_{\alpha}\dot{y}^{\p \alpha}-H(y^{\p}) \label{s2}
\end{equation}
together with  by the primary constraints $\Phi_l (y^{\p})$. In
other words, without losing the generality one can assume that one
is at first given the first order Lagrangian (\ref{a1}) with a
regular presymplectic two-form (\ref{a3}), and then the set of
primary constraints $\Phi^{(1)}_{\mu} (\mu=1,\cdots ,M)$ are
applied to the system. In this way the system is described by the
Lagrangian
\begin{equation}
L=a_i\dot{y}^i-\lambda^{\mu}\Phi^{(1)}_{\mu}-H(y) \label{s3}
\end{equation}
in the extended space $(y^i,\lambda^{\mu})$. The equations of
motion (\ref{a2}) should be replaced in matrix form  by
\begin{equation}
\left( \begin{array}{c|c}f& 0 \\
\hline 0 & 0 \end{array} \right) \left( \begin{array}{c} \dot{y} \\
\hline \dot{\lambda} \end{array} \right)=
\left( \begin{array}{c} \rond H \\
\hline \Phi^{(1)} \end{array} \right) \label{s4}
\end{equation}
which is equivalent  to Eq. (\ref{a2}) together with the
constraint equations $\Phi^{(1)}_{\mu}=0$ $(\mu=1,\cdots,M)$.

Now one should impose the consistency conditions
$\dot{\Phi}^{(1)}_{\mu}=0$. To do this, one should extend the
space to include new variables $\eta^{\mu}$ and add the term
$\eta^{\mu} \dot{\Phi}^{(1)}_{\mu}$ (or equivalently $
-\dot{\eta}^{\mu} \Phi^{(1)}_{\mu}$) to the Lagrangian (\ref{s3}).
This leads in the extended space $(y,\lambda,\eta)$ to the
equations
\begin{equation}
\left( \begin{array}{c|c|c} f& 0 &  A \\  \hline 0&0&0 \\ \hline
-\tilde{A} & 0 & 0 \end{array} \right) \left( \begin{array}{c}
\dot{y} \\
\hline \dot{\lambda} \\ \hline \dot{\eta} \end{array} \right)=
\left( \begin{array}{c} \rond H \\
\hline \Phi^{(1)} \\ \hline 0 \end{array} \right)
 \label{s6}
\end{equation}
where the elements of the rectangular matrix $A$ are given by
\begin{equation}
A_{\mu i}=\rond _i \Phi^{(1)}_{\mu}. \label{s6.5}
\end{equation}
However, nothing would be lost if one forgets about the variables
$\lambda ^{\mu}$ and reduces the system to the Lagrangian
\begin{equation}
L^{(1)}=a_i \dot{y}^i- \dot{\eta}^{\mu}\Phi ^{(1)}_{\mu}-H(y).
\label{s7}
\end{equation}
This leads to the symplectic two-form
\begin{equation}
F = \left( \begin{array}{c|c}f& A \\
\hline -\tilde{A}  & 0
\end{array} \right)
 \label{s8}
\end{equation}
in the $(2K+M)$ dimensional space of variables
$Y\equiv(y^i,\eta^{\mu})$. It should be noted that the Lagrangian
$L^{(1)}$ in Eq. (\ref{s7}) is the same as Eq. (\ref{s3}) in which
$\lambda^{\mu}$ is replaced by $\dot{\eta}^{\mu}$. This means that
the derivatives $\dot{\eta}^{\mu}$ have the same role as
Lagrangian multipliers $\lambda^{\mu}$ corresponding to primary
constraints in the total Hamiltonian
\begin{equation}
H_T=H+\lambda^{\mu}\Phi^{(1)}_{\mu}.
 \label{s9}
\end{equation}
In other words, if some of $\dot{\eta}^{\mu}$'s are found by the
dynamical equations of the system, then the corresponding Lagrange
multipliers are obtained. In Dirac approach \cite{BGP} this would
be the case if there exist some second class constraints.

The equations of motion due to the Lagrangian $L^{(1)}$ can be
written in matrix notation as
\begin{equation}
F \dot{Y}= \partial H. \label{a8}
\end{equation}
Using operations that keep the determinant invariant, it is easy
to show that
\begin{eqnarray}
\nonumber\det F&=&\det \left( \begin{array}{c|c} f& A \\
\hline 0 & \tilde{A}f^{-1}A
\end{array} \right) \nonumber \\ {} &=&(\det f)(\det \tilde{A} f^{-1}
A). \label{a9}
\end{eqnarray}
Since  $\det{f}\neq 0$, $F$ would be singular if $C \equiv
\tilde{A} f^{-1} A $ is singular. Using (\ref{a4}) and
(\ref{s6.5}) we have
\begin{equation}
C_{\mu \nu}=\left\{\Phi^{(1)}_{\mu},\Phi^{(1)}_{\nu} \right\}.
\label{a10}
\end{equation}
Suppose $\textrm{rank}(C)=M^{\z}$ where $M^{\z} \leq M$. This
means that $F$ possesses $M^{\p}=M-M^{\z}$ null-eigenvectors. One
can, in principle, divide $\Phi^{(1)}_{\mu}$'s in two sets
$\Phi^{(1)}_{\mu^{\p}}$ and $\Phi^{(1)}_{\mu^{\z}}$ such that
\begin{equation}
\begin{array}{cc}
\left\{\Phi^{(1)}_{\mu^{\p}},\Phi^{(1)}_{\nu} \right\}\approx 0
&  \\
\hspace{1cm}\left\{\Phi^{(1)}_{\mu^{\z}},\Phi^{(1)}_{\nu^{\z}}
\right\}\approx C_{\mu^{\z} \nu^{\z}}, &\det{C_{\mu^{\z}\nu^{\z}}}
\neq 0. \label{s10}
\end{array}
\end{equation}
where the weak equality symbol $\approx$ means equality on the
surface of the constraints already known (here, the primary
constraints). The matrix $A$ can be decomposed to $A^{\p}$ and
$A^{\z}$ such that
\begin{equation}
\begin{array}{c}
A_{\mup i}=\rond _i \Phi^{(1)}_{\mup}\\
A_{\muz i}=\rond _i \Phi^{(1)}_{\muz}. \label{s11}
\end{array}
\end{equation}
Accordingly  the symplectic tensor $F$ can be written as
\begin{equation}
F=\left( \begin{array}{c|c|c} f& A^{\z} &  A^{\p} \\  \hline
-\tilde{A}^{\z}&0&0 \\
\hline -\tilde{A}^{\p} & 0 & 0 \end{array} \right).
 \label{s12}
\end{equation}
Consider the rectangular matrix
\begin{equation}
\left( \tilde{A}^{\p}f^{-1},0,1\right) \label{s13}
\end{equation}
which has $M^{\p}$ rows and $2K+M$ columns. Using (\ref{s10}) one
can show that its rows are left null-eigenvectors of $F$.
Multiplying (\ref{s13}) with the equations of motion (\ref{a8})
gives the second level constraints as
\begin{equation}
\Phi^{(2)} _{\mup} \approx \left\{ \Phi^{(1)}_{\mup},H \right\}=0.
\label{s14}
\end{equation}
On the other hand, $F$ in (\ref{s12}) has an invertible sub-block
\begin{equation}
F_{\textrm{inv}}=\left( \begin{array}{c|c} f& A^{\z} \\  \hline
-\tilde{A}^{\z}&0 \end{array} \right)
 \label{s15}
\end{equation}
with the inverse
\begin{equation}
F_{\textrm{inv}}^{-1} = \left( \begin{array}{c|c}f^{-1}-f^{-1}
A^{\z}
C^{\z -1} \tilde{A}^{\z} f^{-1} & -f^{-1} A^{\z} C^{\z -1} \\
\hline C^{\z -1} \tilde{A}^{\z} f^{-1} & C^{\z -1} \end{array}
\right). \label{s16}
\end{equation}
This can solve the equations of motion (\ref{a8}) for variables
$\dot{\eta}^{\muz}$ to give
\begin{equation}
\dot{\eta}^{\muz} =-C^{\muz \nuz} \left\{ \Phi^{(1)}_{\nuz},H
\right\} \label{s17}
\end{equation}
where $C^{\muz \nuz}$ is the inverse of $C_{\muz \nuz}$. Inserting
this in the Lagrangian (\ref{s7}) gives
\begin{equation}
L^{(1)}=a_i(y) \dot{y}^i- \dot{\eta}^{\mup}\Phi
^{(1)}_{\mup}-H^{(1)}(y) \label{s18}
\end{equation}
where
\begin{equation}
H^{(1)}= H- \left\{ H, \Phi^{(1)}_{\muz}\right\}C^{\muz \nuz}
\Phi^{(1)}_{\nuz}. \label{s19}
\end{equation}

In this way a number of Lagrange multipliers corresponding to the
second class constraints are derived whose effect is only
replacing the canonical Hamiltonian $H$ with $ H^{(1)}$. Now we
can forget about them and suppose that we are given the primary
constraints $\Phi^{(1)}_{\mu}$and the second level constraints
$\phi^{(2)}_{\mu}$. Next, we should consider the consistency of
$\Phi^{(2)}_{\mu}$ and add the term $-\dot{\eta}^{\mu}_2
\Phi^{(2)}_{\mu}$ to the Lagrangian $L^{(1)}$. Renaming  the
previous $\eta^{\mup}$'s as $\eta^{\mu}_1 $, the new Lagrangian
would be
\begin{equation}
L^{(2)}=a_i(y) \dot{y}^i- \dot{\eta}^{\mu}_1\Phi ^{(1)}_{\mu}
-\dot{\eta}^{\mu}_2\Phi ^{(2)}_{\mu} -H^{(1)}(y) \label{s20}
\end{equation}
this gives the symplectic two-form
\begin{equation}
F^{(2)}=\left( \begin{array}{c|c|c} f& A^{(1)} &  A^{(2)} \\
\hline
-\tilde{A}^{(1)}&0&0 \\
\hline -\tilde{A}^{(2)} & 0 & 0 \end{array} \right)
 \label{s21}
\end{equation}
in the space $\left(y,\eta_1,\eta_2 \right) $. Assuming that the
composed matrix $A\equiv\left( A^{(0)},A^{(1)}\right)$, $F^{(2)}$
has the same from as (\ref{s8}). One should again proceed in the
same way to find the null-eigenvectors
as well as the invertible
sub-block of $F^{(2)}$. The process goes on in this and the
subsequent steps as explained in more detail in \cite{shimoj}.

The important point to be emphasized is that the Lagrangian
\begin{equation}
L^{(n)}=a_i(y) \dot{y}^i- \sum_{k=1}^{n} \dot{\eta}^{\mu}_k\Phi
^{(k)}_{\mu}
 -H^{(n)}(y) \label{s22}
\end{equation}
at the n-th level, say, is equivalent to a system with extended
Hamiltonian
\begin{equation}
H^{(n)}_E=H^{(n-1)}+\sum_{k=1}^{n}\lambda^{\mu}_k \Phi^{(k)}_{\mu}
\label{s23}
\end{equation}
at that level. In other words, the symplectic analysis is
equivalent to the Dirac approach in the context of level by level
method provided that at each level one adds the new constraints
with the corresponding Lagrange multipliers to the Hamiltonian. In
fact this slight difference with the standard Dirac method may
lead to some difficulties as we will see in the following section.

\section{The problem with extended Hamiltonian}

The extended Hamiltonian formalism is well-known in the context of
first class constraints \cite{hendirac,BGP}. In fact, it can be
shown that the dynamical equation
\begin{equation}
\dot{g}=\left\{g,H_E \right\}, \label{s24}
\end{equation}
leads to the correct equation of motion provided that $g$ is a
gauge invariant quantity. In Eq. (\ref{s24}) the extended
Hamiltonian $H_E$ is defined as
\begin{equation}
H_E=H+ \lambda^m \Phi_m \label{s25}
\end{equation}
where $\Phi_m$ are only first class constraints (primary or
secondary). For a first class system, the extended Hamiltonian can
also be used step by step during the process of producing the
constraints . In other words, when all of the constraints are
first class, there is no difference whether one uses
$\dot{\Phi}=\left\{\Phi,H_T \right\}$ or
$\dot{\Phi}=\left\{\Phi,H_E \right\}$.

Now we show that the extended Hamiltonian formalism in Dirac
approach is not suitable when second class constraints are
present. We show this point for a system with only one primary
constraint, i. e. a one-chain system in the language of chain by
chain method. We remember that for such a system level by level
and chain by chain methods coincide.

Consider a system with the canonical Hamiltonian $H(y)$ and one
primary constraint $\Phi^{(1)}$. The total Hamiltonian reads
\begin{equation}
H_T=H+\lambda \Phi^{(1)}. \label{s26}
\end{equation}
Suppose the consistency of $\Phi^{(1)}$ leads to
$\Phi^{(2)}=\left\{ \Phi^{(1)},H \right\}$. Then $\Phi^{(3)}$
emerges as $\left\{ \Phi^{(2)},H \right\}$, and so on. The
iterative process that produces the constraints is described by
\begin{equation}
\Phi^{(n+1)}=\left\{ \Phi^{(n)},H \right\}. \label{s27}
\end{equation}
The above procedure progresses unless $\left\{ \Phi^{(N)},H_T
\right\} \approx 0 $ or $\left\{ \Phi^{(N)},\Phi^{(1)} \right\}
\neq 0$ at the last step $N$. In the former  case the constraints
in the chain are first class, i.e. commute with each other
\cite{lorshir}; while in the latter all the constraints are second
class which means that the matrix
\begin{equation}
C^{nm}=\left\{ \Phi^{(n)}, \Phi^{(m)} \right\} \label{s28}
\end{equation}
is invertible. In this case the Lagrange multiplier $\lambda$
would finally be determined as
\begin{equation}
\lambda=\frac {\left\{ \Phi^{(N)},H \right\}}{\left\{ \Phi^{(N)},
\Phi^{(1)} \right\}}.
 \label{s29}
\end{equation}
Using the Jacobi identity, it is shown in \cite{lorshir} that the
matrix $C^{nm}$ in Eq. (\ref{s28}) has the following form
\begin{equation}
C \approx \left( \begin{array}{ccccc} 0 & 0 & \cdots & 0 & C^{1N}
\\ 0 & 0  & \cdots & C^{2 (N-1)}&C^{2N} \\ \vdots & \vdots & { } &
\vdots &\vdots \\
0 & C^{(N-1) 2} & \cdots & C^{(N-1) (N-1)} & C^{(N-1) N} \\
C^{N1} & C^{N2} & \cdots & C^{N (N-1)} & C^{NN}\end{array}
\right). \label{s30}
\end{equation}
In other words
\begin{equation}
\begin{array}{ccc}
\left\{ \Phi^{(i)}, \Phi^{(j)}\right\} \approx 0 & & if
\hspace{1cm} i+j\leq N.
 \label{s31}
 \end{array}
\end{equation}
Moreover using the Jacobi identity  one can show from
(\ref{s27})that
\begin{equation}
\left\{ \Phi^{(1)}, \Phi^{(N)} \right\} \approx -\left\{
\Phi^{(2)}, \Phi^{(N-1)} \right\} \approx \cdots \approx
(-1)^{(\frac{N}{2}-1)}\left\{ \Phi^{(\frac{N}{2})},
\Phi^{(\frac{N}{2}+1)} \right\} \neq 0.
 \label{s32}
\end{equation}
Remember that $N$ is the number of second class constraints and
necessarily should be even.

 Now suppose  that in order to define the dynamics of the system at
some level $n$, one wishes to use the extended
 Hamiltonian
\begin{equation}
H^{(n)}_E=H+\sum _{k=1}^n \lambda_k \Phi^{(k)}. \label{s33}
\end{equation}
If $n\leq \frac{N}{2}$ then from (\ref{s30}) the consistency of
the constraint $\Phi^{(n)}$ gives
\begin{equation}
\dot{\Phi}^{(n)}=\left\{ \Phi^{(n)},H^{(n+1)}_E \right\} \approx
\left\{ \Phi^{(n)}, H \right\}
 \label{s34}
\end{equation}
which by (\ref{s27}), is the same as $\Phi^{(n+1)}$. However at
level $\frac{N}{2}+1$ the consistency of $\Phi^{(\frac{N}{2}+1)}$,
using $H_E^{(\frac{N}{2}+1)}$ gives
\begin{equation}
\dot{\Phi}^{(\frac{N}{2}+1)}=\left\{ \Phi^{(\frac{N}{2}+1)},H
\right\} +\lambda_{\frac{N}{2}} \left\{ \Phi^{(\frac{N}{2}+1)},
\Phi^{(\frac{N}{2})} \right\}.  \label{s35}
\end{equation}
As is apparent from (\ref{s32}) the above equation solves the
Lagrange multiplier $\lambda_{ \frac{N}{2}}$. There is no
justification to keep $\left\{ \Phi^{(\frac{N}{2}+1)},H \right\} $
as the next constraint $ \Phi^{(\frac{N}{2}+2)}$. In order to knit
the second class chain up to the last element $\Phi^{(N)}$, one is
just allowed to use the total Hamiltonian (\ref{s26}). In other
words, the second half of the chain can be derived if only the
primary constraint $\Phi^{(1)}$ is present in the corresponding
Hamiltonian. As explained in the previous section, using the
standard symplectic analysis  is equivalent to using the extended
Hamiltonian formalism described above. So one should expect some
contradiction in symplectic analysis when second class constraints
are present. In the next section we will show the essence of this
contradiction  for a one chain system and propose a method to
resolve it.
\section{Second class one -chain in symplectic analysis}
According to the algorithm given in section 2, given the canonical
Hamiltonian $H(y)$ and the primary constraint $\Phi^{(1)}_{\mu}$,
at the first step of consistency one should consider the
Lagrangian (see \ref{s7})
\begin{equation}
L^{(1)}=a_i \dot{y}^i- \dot{\eta}_1 \Phi ^{(1)}-H(y). \label{s36}
\end{equation}
The equations of motion can be written in matrix form as
\begin{equation}
\left( \begin{array}{c|c}f& A^{(1)} \\
\hline - \tilde{A}^{(1)} & 0 \end{array} \right) \left(
\begin{array}{c} \dot{y} \\
\hline \dot{\eta}_1 \end{array} \right)=
\left( \begin{array}{c} \rond H \\
\hline 0 \end{array} \right). \label{s37}
\end{equation}
It is easy to see that
\begin{equation}
u^1 \equiv \left( \tilde{A}^{(1)}f^{-1},1\right) \label{s38}
\end{equation}
is the null-eigenvector of the matrix
\begin{equation}
F=\left( \begin{array}{c|c}f& A^{(1)} \\
\hline - \tilde{A}^{(1)} & 0 \end{array} \right). \label{s38.5}
\end{equation}
Implying $u^1$ on both sides of (\ref{s37}) and using (\ref{a4})
gives the new constraint
\begin{equation}
\Phi^{(2)}=\left\{ \Phi^{(1)},H \right\}. \label{s38.1}
\end{equation}
Adding the term $- \dot{\eta}_2 \Phi^{(2)}$ to the Lagrangian (to
perform consistency) gives
\begin{equation}
L^{(2)}=a_i \dot{y}^i- \dot{\eta}_1 \Phi ^{(1)}- \dot{\eta}_2 \Phi
^{(2)}-H(y). \label{s38.2}
\end{equation}
The equations of motion are
\begin{equation}
\left( \begin{array}{c|c|c}f& A^{(1)} & A^{(2)}\\
\hline - \tilde{A}^{(1)} & 0 & 0 \\
\hline - \tilde{A}^{(2)} & 0 &0
 \end{array} \right) \left( \begin{array}{c} \dot{y} \\
\hline \dot{\eta}_1 \\ \hline \dot{\eta}_2
\end{array} \right)=
\left( \begin{array}{c} \rond H \\
\hline 0 \\ \hline 0 \end{array} \right) \label{s39}
\end{equation}
Assuming $\left\{  \Phi^{(1)}, \Phi^{(2)}\right\} \approx 0$, one
can find the new null eigenvector
\begin{equation}
u^2 \equiv \left( \tilde{A}^{(2)}f^{-1},0,1\right). \label{s40}
\end{equation}
Multiplying $u^2$  by (\ref{s39}) gives the new constraint
$\Phi^{(3)}=\left\{ \Phi^{(2)},H \right\}$, and so on.

Suppose one wishes to proceed in this way to find the constraints
of the chain discussed in the previous section, i.e. the second
class chain $\Phi^{(1)},\cdots,\Phi^{(N)}$ with the algebra given
in (\ref{s30}-\ref{s32}). Suppose the above procedure has been
proceeded up to  the step $\frac{N}{2}+1$ where the equations of
motion are
\begin{equation}
\left( \begin{array}{c|c|c|c}f& A^{(1)} & \cdots &
A^{(\frac{N}{2}+1)}\\
\hline - \tilde{A}^{(1)} & 0 & \cdots & 0 \\
\hline \vdots &\vdots&\vdots&\vdots \\
 \hline - \tilde{A}^{(\frac{N}{2}+1)} & 0&\cdots &0
 \end{array} \right) \left( \begin{array}{c} \dot{y} \\
\hline \dot{\eta}_1 \\ \hline \vdots \\ \hline
\dot{\eta}_{\frac{N}{2}+1}
\end{array} \right)=
\left( \begin{array}{c} \rond H \\
\hline 0 \\ \hline \vdots \\ \hline 0 \end{array} \right).
\label{s41}
\end{equation}

Clearly no more null-eigenvector can be find. In fact adding the
column and row corresponding to the constraint
$\Phi^{(\frac{N}{2}+1)}$ has increased  the rank of the matrix $F$
by two. This means that the equations of motion can be solved to
find $\dot{\eta}^{(\frac{N}{2})}$ and
$\dot{\eta}^{(\frac{N}{2}+1)}$. There is no way in the context of
symplectic analysis to proceed further to find the remaining
constraints $\Phi^{(\frac{N}{2}+2)}, \cdots , \Phi^{(N)}$ of the
chain. This is really the failure of traditional symplectic
analysis. In fact this is the reason why the symplectic analysis
has failed in the example given in \cite{rothe} (Particle in hyper
sphere). We will discuss this example in section(6).

What we showed here is the failure of symplectic analysis for a
second class system with only one primary constraint (i.e. a one
chain system). However, one can easily observe that for an
arbitrary system with several primary constraints again the
symplectic analysis would  fail. The reason is that for such a
system some of the constraints driven at level $n$, i.e.
$\Phi^{(n)}_{\mu}$, may have non vanishing Poisson brackets with
constraints of previous levels while  commuting with primary
constraints. As we know from Dirac approach, in such a case the
Poisson brackets of these constraints with Hamiltonian give the
next level constraints. Meanwhile, a little care on symplectic
analysis shows that in this case a number of Lagrange Multipliers
corresponding to non-primary constraints would be determined and
there is no way to find the next level constraints. In this way,
we conclude that {\it the symplectic analysis would fail whenever
second class constraints emerge at third level  or higher}.

\section{How to solve the problem}

In this section we try to find a way to maintain the symplectic
analysis by imposing some modifications. The origin of the problem
is the fact that $ \Phi^{(\frac{N}{2}+1)}$ has non-vanishing
Poisson bracket with $ \Phi^{(\frac{N}{2})}$. As a result, the
symplectic two-form on the left hand side of Eq. (\ref{s41}), i.e.
\begin{equation}
F=\left( \begin{array}{c|c|c|c}f& A^{(1)} & \cdots &
A^{(\frac{N}{2}+1)}\\
\hline - \tilde{A}^{(1)} & 0 & \cdots & 0 \\
\hline \vdots &\vdots&\vdots&\vdots \\
 \hline - \tilde{A}^{(\frac{N}{2}+1)} & 0&\cdots &0
 \end{array} \right),
\label{s42}
\end{equation}
does not possess a new null-eigenvector. If one could consider the
vector
\begin{equation}
u^{(\frac{N}{2}+1)} \equiv \left(
\tilde{A}^{(\frac{N}{2}+1)}f^{-1},0,,\cdots,0,1\right),
\label{s43}
\end{equation}
as a null-eigenvector, then by multiplying $ u^{(\frac{N}{2}+1)}$
on the right hand side of (\ref{s41}), one would obtain the next
constraint as
\begin{equation}
\Phi^{(\frac{N}{2}+2)}=\left\{ \Phi^{(\frac{N}{2}+1)},H \right\}.
\label{s44}
\end{equation}

To reach this goal one should truncate those columns of $F$ which
are located after $A^{(1)}$. In other words, instead of $F$ in Eq.
(\ref{s42}) one should consider the rectangular matrix
\begin{equation}
\tilde{F}=\left( \begin{array}{c|c}f& A^{(1)} \\
\hline - \tilde{A}^{(1)} & 0 \\
\hline \vdots &\vdots \\
 \hline - \tilde{A}^{(\frac{N}{2}+1)} & 0
 \end{array} \right).
\label{s45}
\end{equation}
Clearly $ u^{(\frac{N}{2}+1)}$ in Eq. (\ref{s43}) is the
null-eigenvector of $\tilde{F}$. It is obvious that if one does
the same thing in the subsequent steps, one can produce all the
remaining constraints of the chain, i.e.
$\Phi^{(\frac{N}{2}+1)},\cdots,\Phi^{(N)} $. In the last step the
chain terminates, since $\left\{ \Phi^{(N)},\Phi^{(1)}\right\}
\neq 0 $.

But what is the justification to find the null-eigenvectors of
$\tilde{F}$, i.e. the \textit{truncated} $F$. In fact using Eq.
(\ref{a4}) the set of equations
\begin{equation}
\left( \begin{array}{c|c}f& A^{(1)} \\
\hline - \tilde{A}^{(1)} & 0  \\
\hline \vdots &\vdots \\
 \hline - \tilde{A}^{(N)} & 0
 \end{array} \right) \left( \begin{array}{c} \dot{y} \\
\hline \dot{\eta}_1 \end{array} \right)=
\left( \begin{array}{c} \rond H \\
\hline 0 \\ \hline \vdots \\ \hline 0 \end{array} \right).
\label{s46}
\end{equation}
is equivalent to
\begin{equation}
\begin{array}{ccc}
\hspace{2.3cm} \dot{y}_i=\left\{
y_i,H+\dot{\eta}_1\Phi^{(1)}\right\}&& i=1, \cdots ,2K
\\
\dot{\Phi}^{(j)}=0 & &j=1, \cdots ,N.
\end{array}
\label{s47}
\end{equation}
Remembering that $\dot{\eta}_1$ has the same role as the Lagrange
multiplier $\lambda_1$ corresponding to the primary constraint
$\Phi^{(1)}$, we see that Eq. (\ref{s47}) is the correct equation
of motion
\begin{equation}
\dot{y}_i=\left\{ y_i,H _T\right\}.\label{s49}
\end{equation}
On the other hand, it is easy to see that the equations of motion
resulting from Eq. (\ref{s41}) can be written as
\begin{equation}
\dot{y}_i=\left\{ y_i,H_E \right\} \label{s50}
\end{equation}
where $H_E $ contains all derived constraint(including second
class ones).  In fact as we explained before, the correct
equations of motion are (\ref{s47}) and not (\ref{s50}).

Therefore, if one wishes to proceed in the context of symplectic
analysis, one should consider Eq.(\ref{s46}) instead of
Eq.(\ref{s41})

\section{Example} Consider the Lagrangian
\begin{equation}
L=\frac{1}{2}\dot{\bq}^2+ v \left(\bq^2-1 \right)\label{t1}
\end{equation}
where $\bq \equiv \left(q_1,\cdots, q_n  \right)$. The primary
constraint is $P_v$. The corresponding Hamiltonian is
\begin{equation}
H=\frac{1}{2}\bp^2-v\left(\bq^2-1\right)\label{t2}
\end{equation}
where $\bp \equiv \left(p_1,\cdots,p_n \right)$. In the usual
Dirac approach, using the total Hamiltonian $H_T=H+\lambda P_v$,
the consistency of $\Phi^{(1)}=P_v$ gives the following chain of
constraints
\begin{equation}
\begin{array}{l}
\Phi^{(1)}=P_v  \\
\Phi^{(2)}=\bq^2-1 \\ \Phi^{(3)}=2\bq.\bp \\
\Phi^{(4)}=2\left(\bp^2+2v \bq^2 \right) \label{t3}
\end{array}
\end{equation}
As is apparent, $\Phi^{(4)}$ and $\Phi^{(3)}$ are conjugate to
$\Phi^{(1)}$ and  $\Phi^{(2)}$ respectively. It is worth
remembering that although $\Phi^{(3)}$ is second class, when
reaching at third level, the process of consistency should not
stop, i. e. it should be proceeded one level more to find
$\Phi^{(4)}$ which is conjugate to the primary constraint
$\Phi^{(1)}$. In the symplectic approach the corresponding first
order Lagrangian is
\begin{equation}
L=\bp \dot{\bq}+P_v \dot{v}-\frac{1}{2}\bp^2+v\left(\bq^2-1\right)
 -\lambda P_v. \label{t4}
\end{equation}
This gives the singular presymplectic  tensor
\begin{equation}
F= \left( \begin{array}{c|c}f &0 \\ \hline 0& 0 \end{array}
\right) \label{t5}
\end{equation}
 where $f$ is the usual $(2n+2) \times (2n+2)$ symplectic tensor:
\begin{equation}
f= \left( \begin{array}{c|c}0 &-1 \\ \hline 1& 0 \end{array}
\right). \label{t6}
\end{equation}
The equations of motion for $y^i=\left(\bq,v,\bp,P_v,\lambda
\right)$ are $f_{ij}\dot{y}^j=\rond_i H_T$ where $H_T=H+\lambda
P_v $. Clearly this gives the canonical equation of motion with
Hamiltonian $H_T$, together with the constraint equation $P_v=0$.
Adding the consistency term $-\dot{\eta}_1 P_v $ to the Lagrangian
(\ref{t4}), where $\eta_1$ is a new variable  and forgetting about
the term proportional to $\lambda $ (which just reproduces the
primary constraint) one finds
\begin{equation}
L^{(1)}=\bp \dot{\bq}+P_v \dot{v} -\dot{\eta}_1
P_v-\frac{1}{2}\bp^2 +v\left(\bq^2-1\right). \label{t7}
\end{equation}
This gives the equations of motion
\begin{equation}
F^{(1)}_{ij}\dot{Y}^j=\rond_i H \label{t8}
\end{equation}
where $Y^i \equiv  \left(\bq,v,\bp,P_v, \eta_1 \right)$. In the
matrix form we have
\begin{equation}
F^{(1)}=\left( \begin{array}{c|c}f& A^{(1)} \\
\hline - \tilde{A}^{(1)} & 0 \end{array} \right)  \label{t9}
\end{equation}
where $\tilde{A}^{(1)}=\left( \bz,0,\bz,1\right)$. Here, bold
zero$(\bz)$ means a row vector with $n$ zero components.  Clearly
$u^{(1)}=\left( \bz,-1,\bz,0,1\right)$ is the left
null-eigenvector of $F^{(1)}$. Multiplying the equations of motion
(\ref{t8}) from the left by $u^{(1)}$ gives the constraint
$\Phi^{(2)}=\bq^2-1$.

In the next level we have the Lagrangian
\begin{equation}
L^{(2)}=L-\dot{\eta}_1 P_v -\dot{\eta}_2(\bq^2-1) \label{t10}
\end{equation}
written in the space $Y^i \equiv  \left(\bq,v,\bp,P_v,
\eta_1,\eta_2 \right)$. The corresponding symplectic tensor reads
\begin{equation}
F^{(2)}=\left( \begin{array}{c|cc}f& A^{(1)}&A^{(2)} \\
\hline - \tilde{A}^{(1)} & 0 &0\\  - \tilde{A}^{(2)}
&0&0\end{array} \right)  \label{t11}
\end{equation}
where $\tilde{A}^{(2)}=\left( 2\bq,0,\bz,0\right)$. Clearly
$u^{(2)}=\left( \bz,0,2\bq,0,0,1\right)$ is the null-eigenvector
of $F^2$. Multiplying the equations of motion
$F^{(2)}_{ij}\dot{Y}^j=\rond_i H_T$ from the left by $u^{(2)}$
gives the next level constraint $\Phi^{(3)}=2\bq . \bp$. Again
considering another variable $\eta_3$, the third level Lagrangian
would be
\begin{equation}
L^{(3)}=L-\dot{\eta}_1 P_v -\dot{\eta}_2(\bq^2-1)-\dot{\eta}_3(2
\bq .\bp). \label{t12}
\end{equation}
This gives the following symplectic tensor
\begin{equation}
F^{(3)}=\left( \begin{array}{c|ccc}f& A^{(1)}&A^{(2)}&A^{(3)} \\
\hline - \tilde{A}^{(1)} & 0&0&0 \\  - \tilde{A}^{(2)} &0&0&0\\ -
\tilde{A}^{(3)} & 0&0&0
\end{array} \right)  \label{t13}
\end{equation}
where $\tilde{A}^{(3)}=\left( 2\bp,0,2\bq,0\right)$. Now the
crucial point appears. That is, $F^{(3)}$ has no new
null-eigenvector. In fact one expects that multiplying
$u^{(3)}=\left( -2\bq,0,2\bp,0,0,0,1\right)$ by the equations of
motion due to $L^{(3)}$ gives the next constraint
$\Phi^{(4)}=2\left(\bp^2+2v \bq^2 \right)$. However, it can be
easily checked that $u^{(3)}F^{(3)} \neq 0$. Moreover, $u^{(2)}$
(with one additional zero as the last element) is no more the
null-eigenvector of $F^{(3)}$. This means that adding the
$(2n+5)$th row and columns to $F^{(2)}$ has led to increasing the
rank of $F^{(3)}$ by two. In other words, the equations of motion
for $\dot{\eta}_2$ and $\dot{\eta}_3$ can be solved. Unfortunately
without any modification {\bf there is no way to find} the
Lagrangian
\begin{equation}
L^{(4)}=L-\dot{\eta}_1
P_v-\dot{\eta}_2\left(\bq^2-1\right)-\dot{\eta}_3(2\bq .
\bp)-\dot{\eta}_4\left(2\left(\bp^2+2v \bq^2 \right) \right).
 \label{t14}
\end{equation}
If we could find $L^{(4)}$, then we would be able to have
\begin{equation}
F^{(4)}=\left( \begin{array}{c|cccc}f&
A^{(1)}&A^{(2)}&A^{(3)}&A^{(4)} \\
\hline - \tilde{A}^{(1)} & 0&0&0&0 \\  - \tilde{A}^{(2)}
&0&0&0&0\\ - \tilde{A}^{(3)} & 0&0&0&0 \\ - \tilde{A}^{(4)}
&0&0&0&0
\end{array} \right)  \label{t15}
\end{equation}
where $\tilde{A}^{(4)}=\left( 8v\bq,4\bq^2,4\bp,0\right)$. If we
had somehow derived (\ref{t14}) and (\ref{t15}), then the
singularity of symplectic tensor would completely disappear and
$\dot{\eta}_1, \cdots \dot{\eta}_4 $ would be obtained. However,
using the {\it truncated symplectic tensor} at the second step as
\begin{equation}
\tilde{F}^{(2)}=\left( \begin{array}{c|c}f& A^{(1)}\\
\hline - \tilde{A}^{(1)} & 0 \\  - \tilde{A}^{(2)} &0\end{array}
\right)  \label{t16}
\end{equation}
and similarly  $\tilde{F}^{(3)}$ at the third level as
\begin{equation}
\tilde{F}^{(3)}=\left( \begin{array}{c|c}f& A^{(1)} \\
\hline - \tilde{A}^{(1)} & 0 \\  - \tilde{A}^{(2)} &0\\ -
\tilde{A}^{(3)} &0
\end{array} \right)  \label{t17}
\end{equation}
makes it possible to introduce again $u^{(2)}$ and $u^{(3)}$ as
the corresponding left null-eigenvectors of $\tilde{F}^{(2)}$ and
$\tilde{F}^{(3)}$, respectively. This makes us able to find
$\Phi^{(4)}$ as explained before. It should be noted that one can
after all write the complete symplectic tensor $F^{(4)}$.

This example has also been discussed in \cite{rothe}, where some
other reason is proposed as the origin of failure of the
symplectic analysis. The same results as what we derived here can
be found in every second class system possessing at least four
levels of constraints. For example, one can study the simpler
Lagrangian $L=\dot{x}\dot{y} -z(x+y)$ as well as the more
complicated  example of bosonized  Schwinger model in $(1+1)$
dimensions \cite{MitraRaj,JacRaj} given by
\begin{equation}
{\cal{L}} = \frac{1}{2} \partial_\mu \phi \partial^\mu
  \phi + (g^{\mu\nu} - \varepsilon^{\mu\nu}) \partial_\mu\phi
  A_\nu - \frac{1}{4} F_{\mu\nu} F^{\mu\nu} + \frac{1}{2} A_\mu
  A^\mu.
 \label{c1}
 \end{equation}
One can see that the main feature of the above calculations will
more or less appear in all such examples.

{\bf Acknowledgment}

We thank Esmaeil Mosaffa for reading the manuscripts.

\end{document}